\documentclass[referee]{aa}

\def\eg{{\it e.g.\ }}

\def\kms{\ {\rm km\,s^{-1}}}
\def\kpc{\ {\rm kpc\,h_{50}^{-1}}}
\def\gsim{\lower.5ex\hbox{$\; \buildrel > \over \sim \;$}}
\def\lsim{\lower.5ex\hbox{$\; \buildrel < \over \sim \;$}}
\def\0{{$\clubsuit$}}

\usepackage{graphics}
\begin{document}
\thesaurus{07.19.1}
\title{The cluster of galaxies Abell 970.
\thanks{based on observations made at ESO, La Silla (Chile), Haute-Provence
and Pic du Midi Observatories (France)}}
\author{L.~Sodr\'e Jr.\inst{1} \and D.~Proust\inst{2}
\and H. V. Capelato\inst{3} 
\and G. B. Lima Neto\inst{1}
\and H. Cuevas\inst{4}
\and H. Quintana\inst{5}
\and P. Fouqu\'e\inst{6}} 
\offprints{D.~Proust}
\institute{Departamento de Astronomia do IAG/USP, Av. Miguel Stefano 4200,
04301-904, S\~ao Paulo, Brazil  \\
email: laerte@iagusp.usp.br, gastao@iagusp.usp.br
\and
Observatoire de Paris - Section de Meudon, DAEC, F92195 MEUDON CEDEX, France \\
email: Dominique.Proust@obspm.fr
\and
Divis\~ao de Astrof\'{i}sica INPE/MCT, 12225-010, Sao Jos\'e dos Campos, 
Brazil \\
email: hugo@das.inpe.br
\and
Departamento de F\'{\i}sica, Universidad de La Serena, Benavente 980, La Serena,
Chile \\
email: hcuevas@ns.dfuls.cl
\and
Departamento de Astronomia y Astrof\'{\i}sica, Pontificia Universidad
Cat\'olica de Chile, Casilla 104, Santiago 22, Chile \\
email: hquintan@astro.puc.cl
\and
European Southern Observatory, Casilla 19001, Santiago 19, Chile \\
email: pfouque@eso.org}
\date{Received date; accepted date}
\maketitle
\begin{abstract}
     We present a dynamical analysis of the galaxy cluster Abell 970
based on a new set of radial velocities measured
at ESO, Pic du Midi and Haute-Provence observatories. Our analysis
indicates that this cluster has a substructure and is out of dynamical
equilibrium. This conclusion is also supported by differences in the
positions of the peaks of the surface density distribution and X-ray 
emission, as well as by the evidence of a large scale velocity gradient
in the cluster. We also found a discrepancy between the masses inferred with
the virial theorem and with the X-ray emission, what is expected if the
galaxies and the gas inside the cluster are not in hydrostatic equilibrium.
Abell 970 has a modest cooling flow, as is expected if it is 
out of equilibrium as suggested by Allen (1998). We propose that
cooling flows may have an intermittent behavior, with
phases of massive cooling flows being followed by phases without significant 
cooling flows after the acretion of a galaxy group massive enough to disrupt 
the dynamical equilibrium in the center of the clusters. A massive cooling 
flow will be established again, after a new equilibrium is
achieved.

\end{abstract}

\section{Introduction}

In the present accepted paradigm of structure formation, small structures are
the first to collapse, merging then hierarchically to build larger objects.
In this framework,
clusters of galaxies, the largest (nearly) virialized structures, may
be accreting galaxies and/or `dark haloes' even at $z=0$ 
(e.g. Lanzoni et al. 2000). Therefore, the study of galaxy clusters may
offer important pieces of information for  observational cosmology, because
cluster properties depend on cosmological parameters and can be used to
constrain cosmological scenarios. 
For instance, the cluster mass function, which
may be described by the Press--Schechter formalism (Press \& Schechter 1974),
depends on the density parameter $\Omega$ and on the power spectrum amplitude
and shape parameter (e.g., Lacey \& Cole 1994, Bahcall \& Fan 1998). Also, the
morphological and dynamical state of clusters allow us to infer their history
and, again, to constrain cosmological theories of large scale structure 
formation (e.g., Kauffmann et al. 1999).

The mass of a cluster may be estimated by several methods: the 
optical virial mass, from the positions and radial velocities of 
the cluster galaxies; the X-ray mass, from
the X-ray emission of the hot intracluster gas; the gravitational lensing mass,
from the distortions produced on background object images by the gravitational
field of the cluster. However, discrepancy between these estimators are often
found (e.g., Mushotzky et al. 1995, Girardi et al. 1998, Wu et al. 1998). 
Virial mass estimates rely on the assumption of dynamical equilibrium. 
X-ray mass estimates 
also depend on the dynamical equilibrium hypothesis and on the still not well
constrained intra-cluster gas temperature gradient (Irwing et al. 1999).
Finally, mass estimates based on gravitational lensing 
are considered more reliable than the
others (e.g., Mellier 1999) because they are completely independent of the
dynamical status of the cluster, and their discrepancies with other methods may
be due to non-equilibrium effects in the central region of the clusters (Allen
1998).

An important source of departure from equilibrium 
(that may affect mass estimates) are the substructures. Their very 
existence supports the current
view that clusters grow hierarcally by accreting nearby groups and galaxies.
Note that even the frequency and degree of clumpiness in the central regions of
the clusters depends on the cosmology (e.g., Richstone, Loeb \& Turner 1992). 
In many cases substructures are loosely bound and 
can survive only a few crossing time in the
hostile environment of rich clusters. However, they seem to be very common in
present day clusters. A recent estimate by Kolokotronis et al. (2000) indicates
that at least 45\% of rich clusters present optical substructures.
Substructures are detected in both optical and X-ray images in 23\%
of the clusters.
This last number may then be considered a lower limit on the fraction of real
substructures in clusters, and it is large! Indeed, it implies that one in four
clusters may be out of equilibrium due to the presence of a substructure.
The dynamical status of individual clusters should therefore be examined in
detail before being used in other studies.

In this paper we present a study of the dynamical status of the cluster
Abell 970, from the analysis of the positions and velocities of cluster
galaxies, as well as from the intra-cluster gas X-ray emission. Abell 970 has a
richness class $R=0$ and type B-M III (Abell, Corwin \& Olowin 1989). Together
with a few other clusters, (A979, A978 and A993) it is member of the Sextans
supercluster (number 88 in the catalogue of Einasto et al. 1997, and number 378
in the catalogue of Kalinkov, Valtchanov \& Kuneva 1998). It has a moderate
cooling flow (White, Jones \& Forman 1997).

A search in NED database\footnote{The NASA/IPAC Extragalactic Database (NED) is
operated by the Jet Propulsion Laboratory, California Institute of Technology,
under contract with the National Aeronautics and Space Administration} indicates
that only 4 radial velocities are known in the field of the cluster (see Postman
et al. 1992) which, however, have not been published. Here we examine some
properties of the cluster Abell 970, using a set of 69 new radial velocities.
The observations of radial velocities reported here are part of a program to
study the dynamical structure of clusters of galaxies, started some years ago
and with several results already published (see \eg Proust et al. 1987, 1988,
1992, 1995, 2000, Capelato et al. 1991).

This paper is organized as follows. We present in Sect. 2 the details of the
observations and data reduction. In Sect.s 3 and 4 we discuss the galaxy and
the X-ray distributions, respectively. In Sect. 5 we analyze the velocity
distribution of the cluster galaxies. In Sect. 6 we present mass estimates for
the central region of the cluster, derived from the optical and X-ray
observations. In Sect. 7 we discuss the dynamical status of Abell 970.
Finally, in Sect. 8 we summarize our conclusions. We adopt here, whenever
necessary, $H_0 = 50~ h_{50}~ \kms$ Mpc$^{-1}$.

\section{Observations and Data Reductions} 

The new velocities presented in this paper have been obtained with the
1.52m ESO telescope at La Silla (Chile),  the 2.0m telescope at Pic du
Midi (France), and with the 1.93m telescope at Haute-Provence
Observatory (France).

Observations with the 1.52m ESO~telescope were carried out in
February~1996. We used the Boller and Chivens spectrograph at the
Cassegrain focus, equipped with a 600~lines/mm grating blazed at
5000~{\AA} and coupled to an RCA CCD  detector ($1024 \times 640$
pixels) with pixel size of 15~$\mu$m.  The dispersion was 172~{\AA}/mm,
providing spectral coverage from 3750 to 5700~{\AA}. The exposure times
ranged between 30 and 60 minutes, according to the magnitude of the
object. During the run, calibration exposures were made before and
after each galaxy observation using an He-Ar source.
 
Observations with the 1.93m Haute-Provence Observatory telescope were
carried out in November 1997, November 1998 and April 2000. We used the
CARELEC spectrograph at the Cassegrain focus, equipped with a
150~lines/mm grating blazed at 5000~{\AA} and coupled to an EEV CCD
detector (2048x1024 pixels) with pixel size of 13.5~$\mu$m. The
dispersion of 260~{\AA}/mm allowed a spectral coverage from 3600 to
7300~{\AA}. Wavelength calibration was done using exposures of Hg-Ne
lamps.

Part of the velocities were obtained during an observing run at the
2.0m Bernard Lyot telescope at Pic du Midi Observatory in January 1997.
Despite the declination of Abell 970, we used the ISARD spectrograph in its
long-slit mode with a dispersion of 233~\AA/mm with the TEK chip
(1024x1024 pixels) of 25$\mu$m, corresponding to 5.8~\AA/pixel.
Typically two exposures of 2700s each were taken for fields across the
cluster. Wavelength calibration was done using Hg-Ne lamps before and
after each exposure.

Data reduction was carried out with IRAF\footnote{IRAF is distributed
by the National Optical Astronomy Observatories, which are operated by
the Association of Universities for Research in Astronomy, Inc., under
cooperative agreement with the National Science Foundation.} using the
{\it longslit} package.  The spectra were rebinned uniformly in log
wavelength, with a scale of 1~{\AA}/bin. Radial velocities were
determined using the cross-correlation technique (Tonry and Davis 1979)
implemented in the RVSAO package (Kurtz et al. 1991, Mink et al. 1995),
with radial velocity standards obtained from observations of late-type
stars and previously well-studied galaxies.

Table~1\footnote{Table~1 is also
available in electronic form at the CDS via anonymous ftp 130.79.128.5.} 
lists positions and heliocentric velocities for 69 individual 
galaxies in the field of the cluster. The entries of the table are:

\begin{enumerate}

\item identification number 

\item right ascension (J2000)

\item declination (J2000)

\item morphological type determined from a visual inspection of the Palomar
sky survey (POSS) images

\item $b_J$ magnitude from the COSMOS/UKST Southern Sky Object Catalogue

\item heliocentric radial velocity with its error  ($\kms$)

\item R-value derived from Tonry \& Davis (1979)

\item telescope and notes- {\bf e}: 1.52m ESO telescope, {\bf o}: 1.93m 
OHP telescope; {\bf p} 2.0m Pic du Midi telescope \\

\end{enumerate}

For the analysis in the next sections is important to estimate the
completeness level of the velocity sample as a function of the magnitude.
To do that (and also to study the cluster galaxy projected distribution)
we extracted from the COSMOS/UKST Southern Sky Object Catalogue,
supplied by the Anglo-Australian Observatory (Drinkwater, Barnes \&
Ellison 1995) a sample of galaxies in the direction of the cluster.
An examination of Table 1 indicates that, with one exception, 
all velocities were measured in a region of $\sim
27^\prime \times 22^\prime$ (in RA and DEC, respectively), 
centered on the position of galaxy number
1 in Table 1, what assured a good level of completeness in the
core of the cluster. Indeed, by comparing the velocity and the COSMOS
catalogues we verify that, within this region, our velocity sample is
$\sim92$\% complete at $b_J^{cosmos}=18$, $\sim75$\%  at
$b_J^{cosmos}=19$, and $\sim51$\% at $b_J^{cosmos}=19.75$. 

\begin{table*}
\caption[]{Heliocentric redshifts for galaxies.}
\begin{flushleft}
\begin{tabular}{llllllll}
\hline
{\bf GALAXY} & {\bf R.A.} & {\bf DEC.} & {\bf TYPE} & ${\bf b_J^{cosmos}}$ &
{\bf VELOCITY} & {\bf R} & {\bf N} \\
         & (2000) & (2000) & &   & $V {\pm {\Delta}V}$  & & \\ 
\hline
      &            &           &        &           &       &    \\
 01 &             10 17 25.7 & -10 41 21 &   E/D  &  16.57 &  17525  52 &  9.72 & e  \\
 02 &             10 17 24.6 & -10 41 22 &   S0   &  18.00 &  16209  37 &  7.14 & e  \\
 03 &             10 17 28.3 & -10 40 59 &  S0/S  &  18.67 &  18424  74 &  5.48 & e  \\
 04 &             10 17 29.7 & -10 40 31 &   S0   &  17.62 &  16270  52 &  9.67 & e  \\
 05 &             10 17 26.3 & -10 41 34 &   S0   &  17.51 &  18357  54 &  6.60 & e  \\
 06 &             10 17 23.9 & -10 42 16 &  S0/S  &  18.41 &  16435  69 &  5.61 & e  \\
 07 &             10 17 27.3 & -10 41 51 &  E/S0  &  18.36 &  18145  53 &  5.89 & e  \\
 08 &             10 17 29.5 & -10 42 17 &    S   &  18.38 &  17404  71 &  7.42 & e  \\
 09 &             10 17 21.7 & -10 42 56 &    S   &  17.96 &  17631  61 &  6.91 & e  \\
 10 &             10 17 24.3 & -10 43 29 &    S   &  17.95 &  16341  81 &  6.76 & e  \\
 11 &             10 17 31.8 & -10 42 59 &  S0/S  &  18.83 &  17769  81 &  5.79 & e  \\
 12 &             10 17 30.0 & -10 43 09 &   S0   &  18.45 &  17327  35 &  5.57 & e  \\
 13 &             10 17 32.9 & -10 43 52 &    S   &  18.26 &  16974  36 &  5.14 & e  \\
 14 &             10 17 29.4 & -10 44 34 &   Sa   &  17.69 &  18909  75 &  5.61 & e  \\
 15 &             10 17 28.0 & -10 44 18 &   S0   &  17.67 &  18030  29 &  8.46 & e  \\
 16 &             10 17 20.7 & -10 44 16 &    S   &  18.56 &  17711  26 &  5.85 & e  \\
 17 &             10 17 15.6 & -10 43 28 &   Sa   &  17.51 &  19503  96 &  2.62 & e  \\
 18 &             10 17 12.9 & -10 42 47 &   S0   &  17.83 &  19450  51 &  9.10 & e  \\
 19 &             10 17 35.6 & -10 39 55 &   Sb   &  17.15 &  18789  24 &  9.41 & e  \\
 20 &             10 17 28.1 & -10 39 27 &   Sa   &  18.16 &  16842  76 &  4.79 & e  \\
 21 &             10 17 27.6 & -10 39 12 &   S0   &  18.63 &  16404  92 &  4.48 & e  \\
 22 &             10 17 21.0 & -10 40 13 &   S0   &  17.18 &  18788  45 &  9.41 & e  \\
 23 &             10 17 23.2 & -10 40 18 &    E   &  18.26 &  19381  40 & 10.12 & e  \\
 24 &             10 17 22.0 & -10 39 45 &    E   &  18.00 &  16847  47 &  6.37 & e  \\
 25 &             10 17 22.5 & -10 39 50 &  E/S0  &  17.09 &  19483  54 &  9.32 & e  \\
 26 &             10 17 25.2 & -10 41 07 &    E   &  18.87 &  17081  93 &  3.15 & e  \\
 27 &             10 17 28.5 & -10 41 13 &    E   &  17.52 &  33586  98 &  2.84 & e  \\
 28 &             10 17 21.9 & -10 43 12 &  E/S0  &  18.71 &  17764  64 &  5.17 & e  \\
 29 &             10 17 14.4 & -10 39 16 &    S   &  18.46 &  18415  94 &  3.44 & e  \\
 30 &             10 17 12.6 & -10 40 05 &  E/S0  &  17.69 &  16533  80 &  5.36 & e  \\
 31 &             10 17 33.6 & -10 38 46 &   SB   &  17.88 &  17779 109 &  4.68 & e  \\
 32 &             10 17 44.6 & -10 39 14 &  S0/S  &  18.09 &  17992  33 &  7.55 & e  \\
 33 &             10 17 32.4 & -10 34 27 &   S0   &  19.01 &  17606  98 &  2.75 & e  \\
 34 &             10 17 30.7 & -10 36 24 &    S   &  17.43 &  16722 134 &  3.55 & e  \\
 35 &             10 17 41.6 & -10 35 46 &  S0/S  &  17.63 &  21637  37 &  8.62 & e  \\
\end{tabular}
\end{flushleft}
\end{table*}
\begin{table*}
\begin{flushleft}
\begin{tabular}{llllllll}
\hline
{\bf GALAXY}  & {\bf R.A.} & {\bf DEC.} & {\bf TYPE} & ${\bf b_J^{cosmos}}$ &
{\bf VELOCITY} & {\bf R} & {\bf N} \\
         & (2000) & (2000) &  &  & $V {\pm {\Delta}V}$  & & \\ 
\hline
 36 &             10 17 51.1 & -10 35 00 &    E   &  17.29 &  16661  62 &  4.42 & e  \\
 37 &             10 17 36.9 & -10 46 01 &    S   &  16.8~$^7$ &  11957  79 &  4.97 & e  \\
 38 &             10 17 36.8 & -10 46 05 &  S0/S  &  16.8~$^7$ &  12221  77 &  4.70 & e1 \\
 39 &             10 16 55.8 & -10 38 51 &   Sa   &  17.83 &  17570  74 &  7.53 & e  \\
 40 &             10 16 58.5 & -10 38 07 &  S0/S  &  16.42 &  17184  44 &  6.33 & e  \\
 41 &             10 16 59.7 & -10 37 39 &    E   &  18.62 &  17873  59 &  6.44 & e  \\
 42 &             10 17 00.6 & -10 37 12 &    E   &  19.16 &  17194  66 &  5.58 & e  \\
 43 &             10 17 03.9 & -10 37 38 &  E/S0  &  18.11 &  17690  52 &  7.14 & e  \\
 44 &             10 17 11.8 & -10 36 07 &  E/S0  &  16.43 &  21998  84 &  8.33 & e2 \\
 45 &             10 16 57.5 & -10 40 16 &    E   &  17.46 &  17903  48 &  9.11 & e  \\
 46 &             10 17 02.0 & -10 39 59 &    E   &  19.10 &  18843  98 &  4.53 & e3 \\
 47 &             10 17 07.6 & -10 45 46 &   S0   &  16.89 &  17202  86 & 15.72 & e4 \\
 48 &             10 17 10.2 & -10 46 26 &  E/S0  &  17.13 &  17475  46 & 12.24 & e5 \\
 49 &             10 16 49.9 & -10 47 23 &   Sb   &  16.91 &  17487  64 &  6.73 & e  \\
 50 &             10 17 00.5 & -10 47 17 &   Sa   &  17.38 &  21166 107 &  3.77 & e  \\
 51 &             10 16 54.1 & -10 43 10 &  S0/S  &  18.45 &  17257  85 &  4.71 & e  \\
 52 &             10 16 53.2 & -10 43 43 &    E   &  18.83 &  17351  85 &  6.70 & e  \\
 53 &             10 17 47.0 & -10 45 19 &    S   &  19.14 &  16771 138 &  2.96 & p  \\
 54 &             10 18 18.0 & -10 46 48 &  E/S0  &  15.95 &  11579  63 &  8.04 & p  \\
 55 &             10 18 15.6 & -10 45 03 &   S0   &  17.69 &  12056  81 &  3.26 & p  \\
 56 &             10 17 51.6 & -10 44 47 &    E   &  19.44 &  16372  89 &  2.53 & p  \\
 57 &             10 17 54.6 & -10 43 47 &   S0   &  19.44 &  45962  93 &  2.61 & p  \\
 58 &             10 18 06.9 & -10 42 43 &    S   &  19.13 &  51814  76 &  2.41 & p  \\
 59 &             10 18 04.0 & -10 41 44 &   SBc  &  19.68 &  48068  73 &  3.01 & p  \\
 60 &             10 17 57.7 & -10 33 48 &    S   &  18.52 &  16228  44 &  4.53 & p  \\
 61 &             10 18 13.0 & -10 34 42 &    S   &  18.47 &  11675  69 &  5.34 & o6 \\
 62 &             10 16 55.2 & -10 30 49 &    S   &  18.16 &  18371 133 &  3.30 & o  \\
 63 &             10 16 54.7 & -10 33 22 &    S   &  17.86 &  17737  75 &  4.47 & o  \\
 64 &             10 16 36.9 & -10 32 38 &   S0   &  17.58 &  17982  82 &  5.62 & o  \\
 65 &             10 16 51.9 & -10 36 15 &    S   &  17.53 &  18255  84 &  5.39 & o  \\
 66 &             10 16 48.8 & -10 39 10 &   S0   &  18.67 &  18293  63 &  6.48 & o  \\
 67 &             10 17 42.0 & -10 45 49 &    E   &  18.60 &  12101 114 &  3.22 & o  \\
 68 &             10 16 37.3 & -10 32 39 &    ?   &  18.77 &  17589 136 &  3.04 & o  \\
 69 &             10 16 41.3 & -10 52 38 &    S   &  17.37 &  17649  42 &  8.74 & o  \\
\hline
\end{tabular}
\end{flushleft}
\smallskip
{\bf emission lines}: 
{\bf 1} [OIII], H${\alpha}$, [SII]$=~12084\kms$; 
{\bf 2} H${\alpha}=~22087\kms$;
{\bf 3} H${\alpha}=~18858\kms$;
{\bf 4} H${\alpha}=~17234\kms$;
{\bf 5} H${\alpha}=~17523\kms$;
{\bf 6} measured on H${\beta}$, [OIII], H${\alpha}$.
{\bf 7} estimated magnitude (see Sect. 4). 
\end{table*}

\section{Galaxy distribution}

We present in Fig. 1 the distribution of galaxies in the direction of
Abell 970, obtained from the COSMOS catalogue, for galaxies brighter than 
$b_J^{cosmos} < 19.75$ (235
objects). The plot is centered on an E/D galaxy (number 1 in Table 1)
and has $45^\prime \times 45^\prime$ (i.e., about 4.6 $\times$ 4.6
h$_{50}^{-1}$ Mpc).  It is interesting to point out that, amongst the
cluster galaxies with known velocities, this is the
second brightest galaxy ($b_J^{cosmos} = 16.57$). 

\vspace{0.3cm}
{\centering
\resizebox*{8cm}{!}{\rotatebox{-90}{\includegraphics{figure1.ps}}}
\par}
{\centering \small {Fig. 1- Galaxies brighter than $b_J^{cosmos} <
19.75$ in the field of Abell 970. The plot is centered on galaxy number 1 in
Table 1. }}

The adaptive kernel
density map (Silverman, 1986) corresponding to this sample is given in
Fig. 2. This figure indicates that the galaxy distribution in the
field of Abell 970 is approximately regular, with the projected 
density peaking at the
position of the E/D galaxy. There is a substructure at NW, near
galaxy number 40 in Table 1. This is the brightest cluster galaxy (considering
only galaxies with known radial velocities); it is classified as
S0/S and has magnitude $b_J^{cosmos} = 16.42$. This figure also
indicates that the cluster radial extension may attain several Mpc.

It is worth noting that all features displayed in this map are significant.
The significance regions are obtained through
a bootstrap resampling procedure applied to the
sample coordinate distribution. This allows the construction of a map of
standard deviations of the projected density. By subtracting the
projected density map from the standard deviation map (multiplied by a given
number, say 3), we define the {\it significance regions} of the projected
density map as those regions for which the resulting subtracted map
is positive.

{\centering
\resizebox*{12cm}{!}{\rotatebox{-90}{\includegraphics{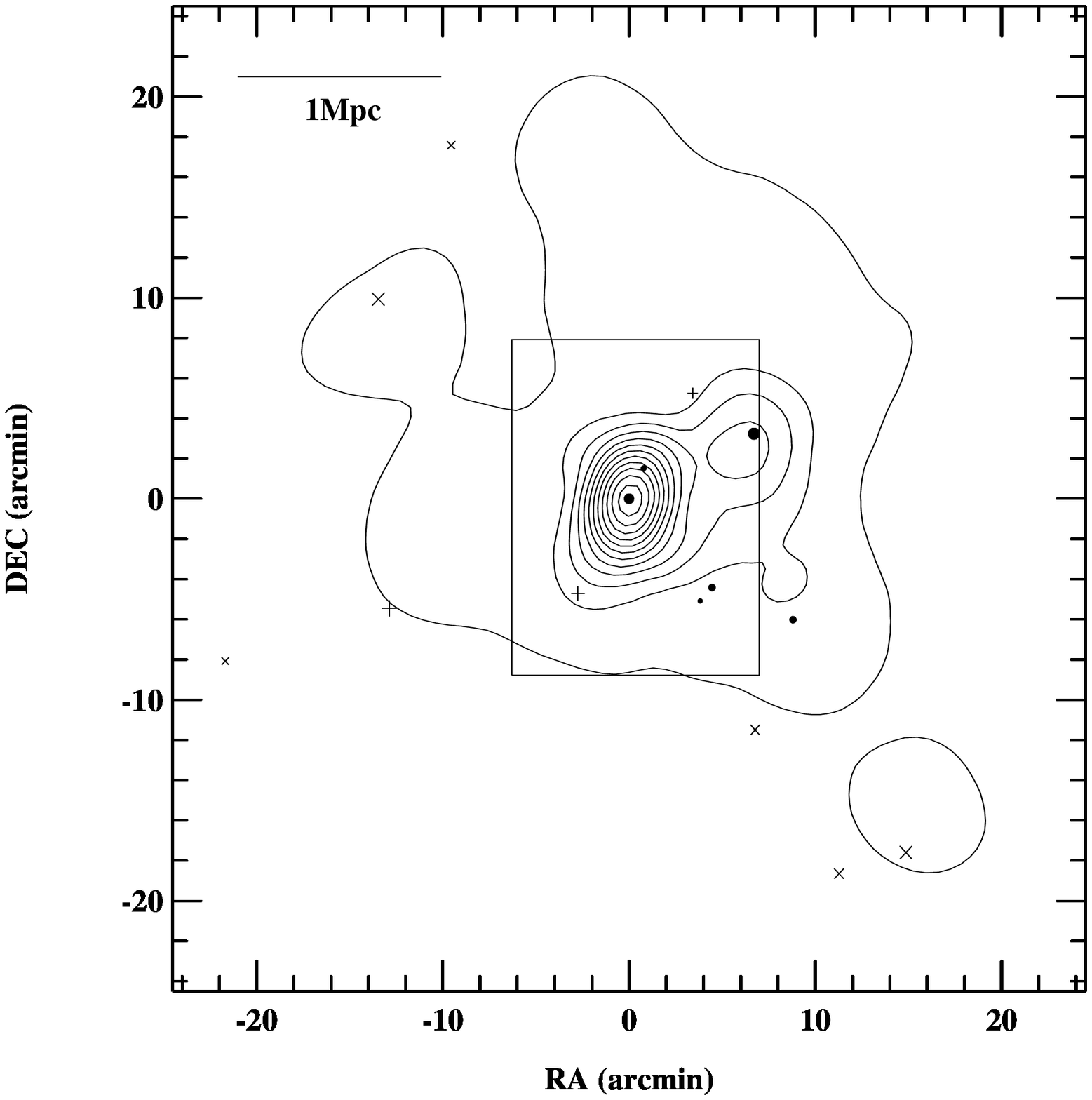}}}
\par}
{\centering \small {Fig. 2- Projected density map of the galaxies brighter 
than $b_J^{cosmos} =19.75$ in the field of Abell 970, together with
the positions of the 15 brightest ones. The peak density of this
map corresponds to about 210 galaxies Mpc$^{-2} h_{50}^2$. 
All density levels in this map are significant.
Other symbols are as follows- Dots:
galaxies kinematically linked to the cluster; Pluses: galaxies
projected in this field but not belonging to the cluster; Crosses:
galaxies with no measured velocities.  The central
rectangle displays the area covered by the X-ray image of Fig. 4.}}
\vspace{0.3cm}

\section{The X-ray emission and the gas distribution}

Abell 970 is a X-ray source, first observed with Einstein IPC in June 1980
(Ulmer et al. 1981). It was also observed during the ROSAT all-sky survey
(Voges 1992), in 1990, and is included in the X-ray-brightest Abell
type cluster catalog (XBACs; Ebeling et al. 1996).

Since Abell 970 had no measured X-ray temperature, Ebeling et al. (1996)
determined the X-ray flux, luminosity and temperature using an
iterative method running roughly as follows:  assuming an initial
X-ray temperature of 5 keV, the bolometric luminosity was computed.
Then, with this luminosity and using the $T_X$--$L_X$ relation
from White, Forman \& Jones (1996), a new estimate of the temperature
was made which, in turn, was used to compute a new luminosity and so
on. Thus, with Rosat data and the above mentioned $T_X$--$L_X$
relation, Ebeling et al. (1996) quoted a flux equal to $9.9 \times
10^{-12}$ erg~cm$^{-2}$~s$^{-1}$, luminosity of $1.50 \times
10^{44}$ erg~s$^{-1}$ (both in the 0.1--2.4 keV band), and gas
temperature $k T_{X} = 3.6$ keV. By applying this same iterative
method  to the Einstein IPC data, Jones \& Forman (1999) derived a X-ray
luminosity of $2.13 \times 10^{44}$ erg~s$^{-1}$ in the [0.5--4.5 keV]
band and a bolometric luminosity equal to  $3.79 \times
10^{44}$ erg~s$^{-1}$, which is in accordance with the temperature given
by Ebeling et al. (1996). Moreover, White, Jones \& Forman (1997),
using this same data, suggests that Abell 970 has a weak cooling-flow, with
a  mass deposition rate $\dot{M} = 20^{+32}_{-20}M_\odot$ yr$^{-1}$
(see also Loken, Mellot \& Miller 1999).

However, since the Einstein IPC detector had some spectroscopic
capability, it is possible to estimate its temperature by a direct
fitting of the available spectra. We have thus obtained both the
`events' and image (in the 0.2--3.5 keV band, rebinned to 24 arcsec per
pixel) files from the HEASARC Online Service. The spectrum of Abell 970  was
extracted from the events file with XSELECT and analysed with XSPEC
using the PI channels 4--12 (0.5--4.5 keV) within a region of 9.6
arcmin (corresponding to 1 $\rm h_{50}^{-1}$~Mpc). The X-ray emission was
fitted with a single temperature, absorbed MEKAL model (Kaastra \& Mewe
1993; Liedahl et al.  1995). We have also used the recipe given by
Churazov et al. (1996) for computing the weights (available in XSPEC),
based on the smoothed observed spectrum. With these weights, one can
still use the least-square minimisation and the $\chi^{2}$ statistics
to estimate the confidence interval of the fitted parameters.

With only 9 bins covering the 0.5--3.5 keV band, it is impossible to
constrain the metallicity. Therefore we have fixed $Z$ to the
`canonical' value of $0.3 ~ Z_\odot$, which is the mean value obtained
for 40 nearby clusters by Fukazawa et al. (1998). Also, with only 3 bins
with energy less than 1~keV, it is difficult to constrain independently
the temperature and the hydrogen column density. This happens because
they are correlated (e.g. Pislar et al. 1997).  Therefore, we have also
fixed the hydrogen column density at $N_{\rm H} = 5.3\times
10^{20}$cm$^{-2}$, the galactic value at the Abell 970 position (Dickey \&
Lockman 1990). Fig. \ref{a85_fit_spec_cstat} shows the fitting of the
X-ray spectrum. The results are summarised in Table \ref{AjusteRaioX}.

\begin{table}[htbp]
\caption[]{X-ray spectral fitting results. $kT$ is the gas temperature, 
$N_{\rm H}$ is the hydrogen column density, $Z$ is the metallicity, $L_X$ 
is the [0.5--4.5 keV] non-absorbed luminosity inside 1 $h^{-1}_{50}$ Mpc, 
and the last column gives the $\chi^2$ and number of degres of freedom. 
Errors are at a confidence level of 68\%.}
\begin{tabular}{c c c c c}
\hline
$kT$   &        $N_{\rm H}$    &     $Z$      &  $L_X$   & $\chi^{2\strut}$/dof \\
(keV)  &  ($10^{20}$cm$^{-2}$) &  ($Z_\odot$) & ($10^{44}$ ergs~s$^{-1}$) \\
\hline
  $3.1_{-0.6}^{+1.0}$ & 5.3$^\star$  &  0.3$^\star$  &  1.8  & 6.57/7 \\
\hline
\label{AjusteRaioX}
\end{tabular}
\smallskip
Notes: $^\star$ Value fixed. Varying the metallicity from 0.1 to 0.5 $Z_\odot$ 
produces a change in temperature of less than $0.3$ keV, increasing towards 
$Z=0.1~ Z_\odot$.
\end{table}

\vspace{0.3cm}
{\centering
\resizebox*{8cm}{!}{\rotatebox{0}{\includegraphics{figure3.ps}}}
\par}
{\centering \small {Fig. 3- Fit of the Abell 970 IPC X-ray spectrum. 
Both the metallicity 
and hydrogen column density are fixed (0.3 $Z_\odot$ and 
$5.3 \times 10^{20}$ cm$^{-2}$, respectively).}}

Although slightly cooler, our temperature is in good agreement with the
one devired by Ebeling et al. (1996) and Jones \& Forman (1999). Using
the $\sigma$--$T_{X}$ relation given by Wu, Xue \& Fang (1999), such a
temperature corresponds to $\sigma$ in the range 640--720~km/s.

In Fig. 4 we display the X-ray isophotes of an Einstein IPC image in the
[0.2--3.5 keV] band. This image has 24 arcsec per pixel. As it can be
seen, the X-ray isophotes are also regular but, interestingly, their
peak is not coincident with the peak of the projected density
distribution, being slightly displaced towards the substructure
associated to the cluster brightest galaxy. Since the relaxation time
of the hot gas is expected to be much lower than that of the galaxies,
the non-coincidence between the peak of the galaxy distribution and the
X-ray emission may be an evidence of a state of
non-equilibrium in the galaxy distribution, as expected if the
substructure associated to the brightest galaxy has only recently
infalled into the cluster. We will explore this point further in next
sections. Note that there is not a X-ray emission excess near the
substructure detected in the projected density distribution. This is
also consistent with the hypothesis that the galaxies in this substructure
are members of a group recently captured by the cluster, whose X-ray emission
is much lower than that of the cluster.

\vspace{0.3cm}
{\centering
\resizebox*{10cm}{10cm}{\rotatebox{0}{\includegraphics{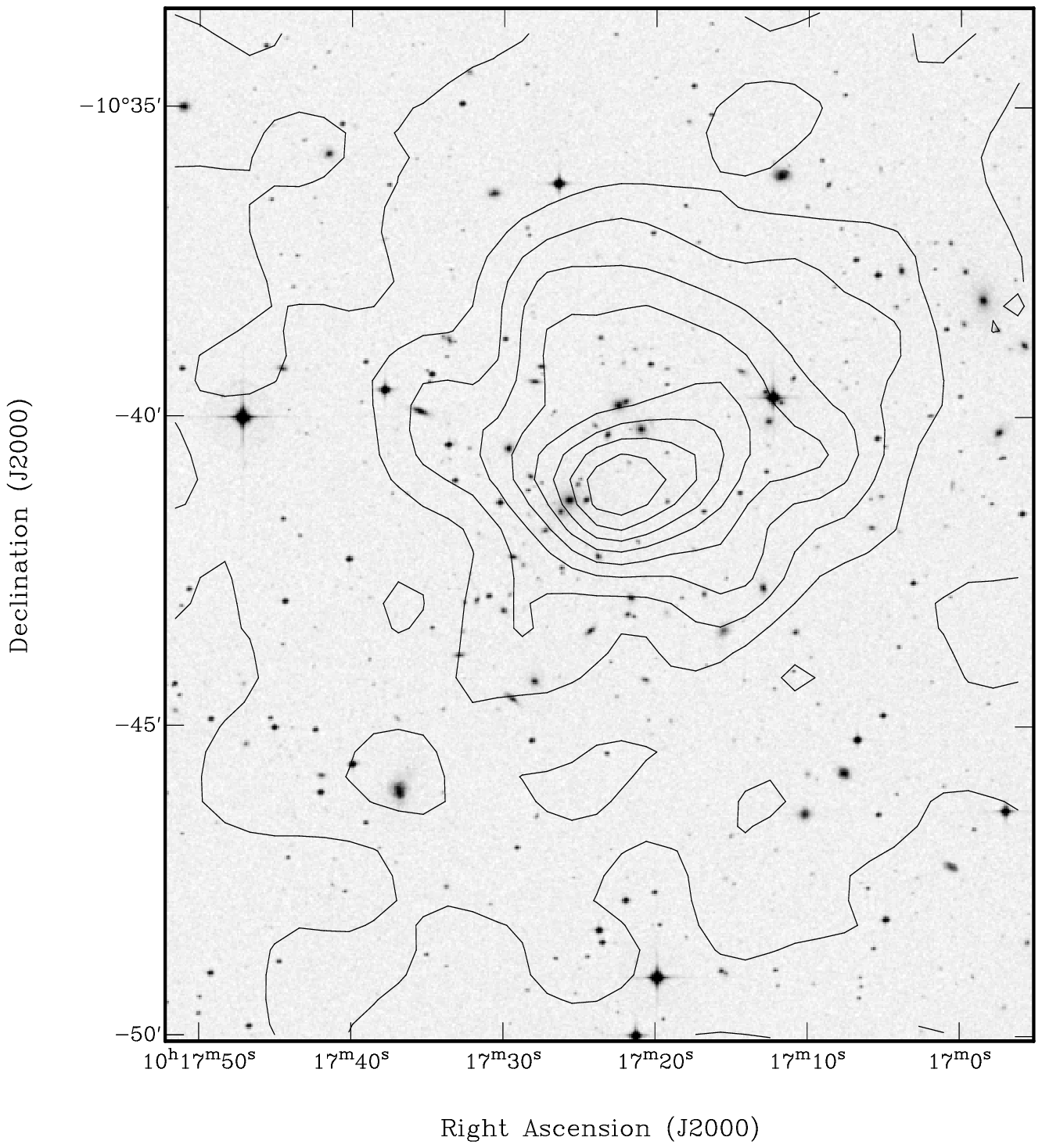}}}
\par}
{\centering \small {Fig. 4- X-ray isophotes from an Einstein IPC
image of Abell 970 (obtained from the HEASARC Online Service) superposed on a DSS image of the cluster.}}
\vspace{0.3cm}

The X-ray isophotes also show  a small peak at SE of the cluster
center, at the position of galaxies number 37 and 38 in Table 1,
 identified in the COSMOS catalogue as just one galaxy with magnitude
$b_J^{cosmos} =16.07$.  However, an examination of the optical image of
this object using POSS indicates that it indeed corresponds to two
merging galaxies. A butterfly-shape due to the tidal currents induced
by the merger can be noticed in the image and the spectrum of object
number 38 has emission lines. It is somewhat surprising that this
system is not cataloged as an IRAS source. The excess of X-ray emission
associated to this galaxy pair may be an evidence of an active nucleus
excited by the merger.  The magnitudes given in Table 1 were estimated
from the COSMOS magnitude assuming that both galaxies have the same
luminosity.

\section{Velocity analysis}
Our sample contains 69 velocities in the direction of Abell 970.
Fig. 5 shows a wedge velocity diagram in the 
direction of the cluster in right ascension (up) and declination (down),
and indicates that most of the velocities are between 15000 and 20000
km s$^{-1}$. A histogram of the velocity distribution of the sample is 
displayed in Fig. 6. In this section we will discuss the velocity 
distribution, looking for non-equilibrium effects.

\vspace{0.3cm}
{\centering \resizebox*{12cm}{4cm}{\rotatebox{-90}{\includegraphics{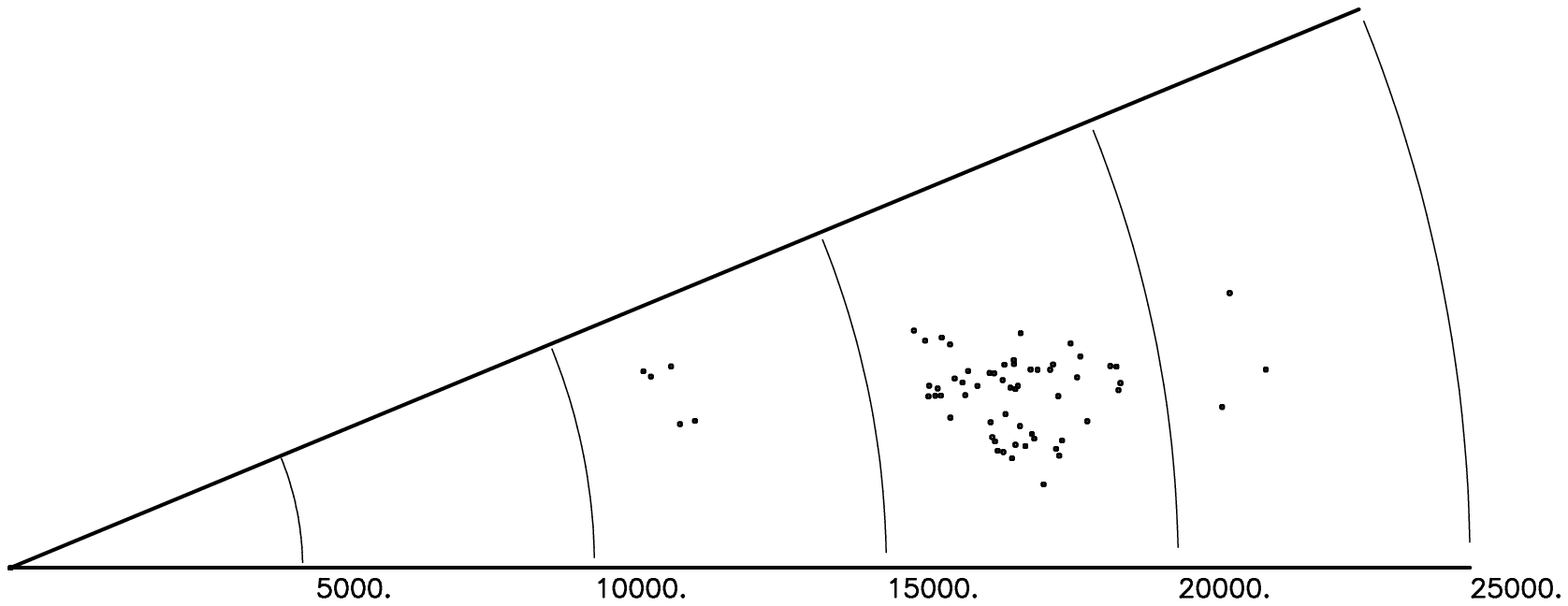}}}
\par}
{\centering \resizebox*{12cm}{4cm}{\rotatebox{-90}{\includegraphics{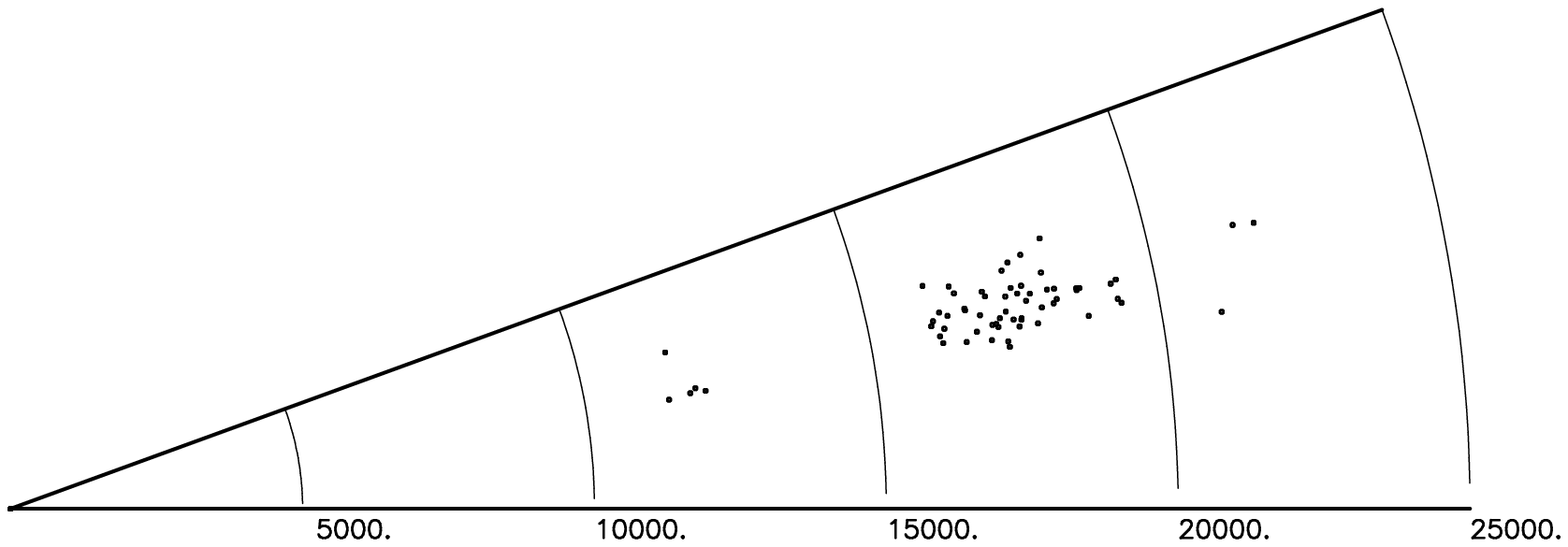}}}
\par}
\vspace{0.3cm}
{\centering \small {Fig. 5- Wedge velocity diagram in right ascension
(up), and declination (down) for the measured galaxies in Abell 970 with
radial velocities smaller than 25000 $\kms$.}}
\vspace{0.3cm}

\subsection{Normality and dispersion of the velocity distribution}

The usual recursive $3\sigma$-clipping (Yahil \& Vidal 1977) fails to
simultaneously remove the low ($v \sim 12000\kms$) and high velocity
($v \sim 22000\kms$) tails of the distribution. We then decided to carefully
analyze the radial velocity distribution from the point of view of
statistical normality tests, as those provided by the ROSTAT package
(Beers et al. 1990; Bird \& Beers 1993). With this aim, we have
examined several different samples of radial velocities:  
sample $A$ comprises all galaxies within $11500\kms < v <
22000\kms$; sample $B$ is similar to sample $A$ without the high- and
low-velocity tail, that is $15000\kms < v < 20000\kms$; sample $D$ is
similar to sample $B$ added with the high-velocity tail, that is
$15000\kms < v < 22000\kms$; sample $C$ contains galaxies within
$15000\kms < v < 18500\kms$ and has been considered in view of a
significant gap in the data occurring at $v \sim 18500\kms$, as
indicated by the gap analysis (Wainer \& Shacht 1978) provided by
ROSTAT.
 
Table \ref{estatistica} lists the values of the statistics and
associated significance levels for the $a$, $u$ and $W$ tests, which are
most sensitive to the tail populations (see discussion in Yahil \&
Vidal 1977) and the Kolmogorof-Smirnov (KS) normality tests. Only
those $p$ values significant at levels better than 10\% are reported.
The other statistical tests provided by ROSTAT have given results
always similar to the KS test and so they have been omitted. The last
two columns give the values of the Asymmetry Index and the Tail Index
(Bird \& Beers 1993), which are robust estimators for the skewness and
kurtosis of the distribution. The Dip test of unimodality of the
distribution also failed to give significant results for any of the
samples.

\begin{table}[htbp]
\caption[]{Statistics of the velocity distribution.}
\tabcolsep=0.6\tabcolsep
\begin{tabular}{c c c c c c c c c c r}
\hline
Sample & N & $a$ & $p(a)$ & $u$ & $p(u)$ & $W$ & $p(W)$ & p(KS) & AI & TI \\
(1) & (2) & (3) & (4) & (5) & (6) & (7) & (8) & (9) & (10) & (11) \\
\hline
$A$ & 65 & 0.639 & 0.01& 5.005 &      & 0.838 & 0.01 & 0.01 &  0.069 & 1.292 \\
$B$ & 56 & 0.780 &     & 3.777 & 0.02 & 0.954 & 0.06 &      &  0.081 & 1.080 \\
$C$ & 59 & 0.711 & 0.01& 4.730 &      & 0.881 & 0.01 & 0.01 &  1.057 & 1.273 \\
$D$ & 48 & 0.829 &     & 3.403 & 0.01 & 0.940 & 0.02 &      & -0.514 & 1.015 \\
\hline
\label{estatistica}
\end{tabular}
\end{table}

The normality hypothesis is rejected for samples
A and C at significance levels better than 3\% for all the statistical
tests, excepting the $u$ test. The values of the statistics of the $a$
and $W$ tests, as well as the TI values, indicate long tailed
underlying distributions.  Removing the tails of $A$ sample, sample
$B$, seems to bring the distribution to normality, as indicated by most
of the tests.  Although both the $u$ and the $W$ tests reject normality
at high significance levels, their results seem contradictory with the
$u$ statistics suggesting a cutoff of the underlying distribution while
the $W$ statistics indicates it is long tailed. Notice that both the AI
and TI values are consistent with a normal underlying distribution.
Very similar results were obtained for sample D, as compared with
sample B, except that in this case the distribution seems significantly
skewed towards low velocities. This is not unexpected for, even if the
marginal indication of bimodality of the distribution  given by the gap
analysis were confirmed, there would be no way, at this level of
analysis, to disentangle galaxies belonging to one or to the other
underlying distributions. Since the Dip statistics also fail to reject
unimodality, we will not consider this possibility for now and will
adopt sample B as representative of the radial velocity distribution of
the cluster. We will return to this question at the end of this
section.

We will thus assume that the cluster galaxies have radial velocities in
the range between $15000$ and $20000\kms$. It is interesting to note that a low
velocity tail at $v \sim 12000\kms$, similar to the one found here,
also affected the velocity distribution of the cluster A979 (Proust et
al. 1995), which is the nearest neighbor cluster of Abell 970 (at about
$3^{\circ}$ NE from its center), within the same supercluster.  This
suggests the existence of a large foreground structure projected in this
region of the sky.

{\centering
\resizebox*{!}{10cm}{\includegraphics{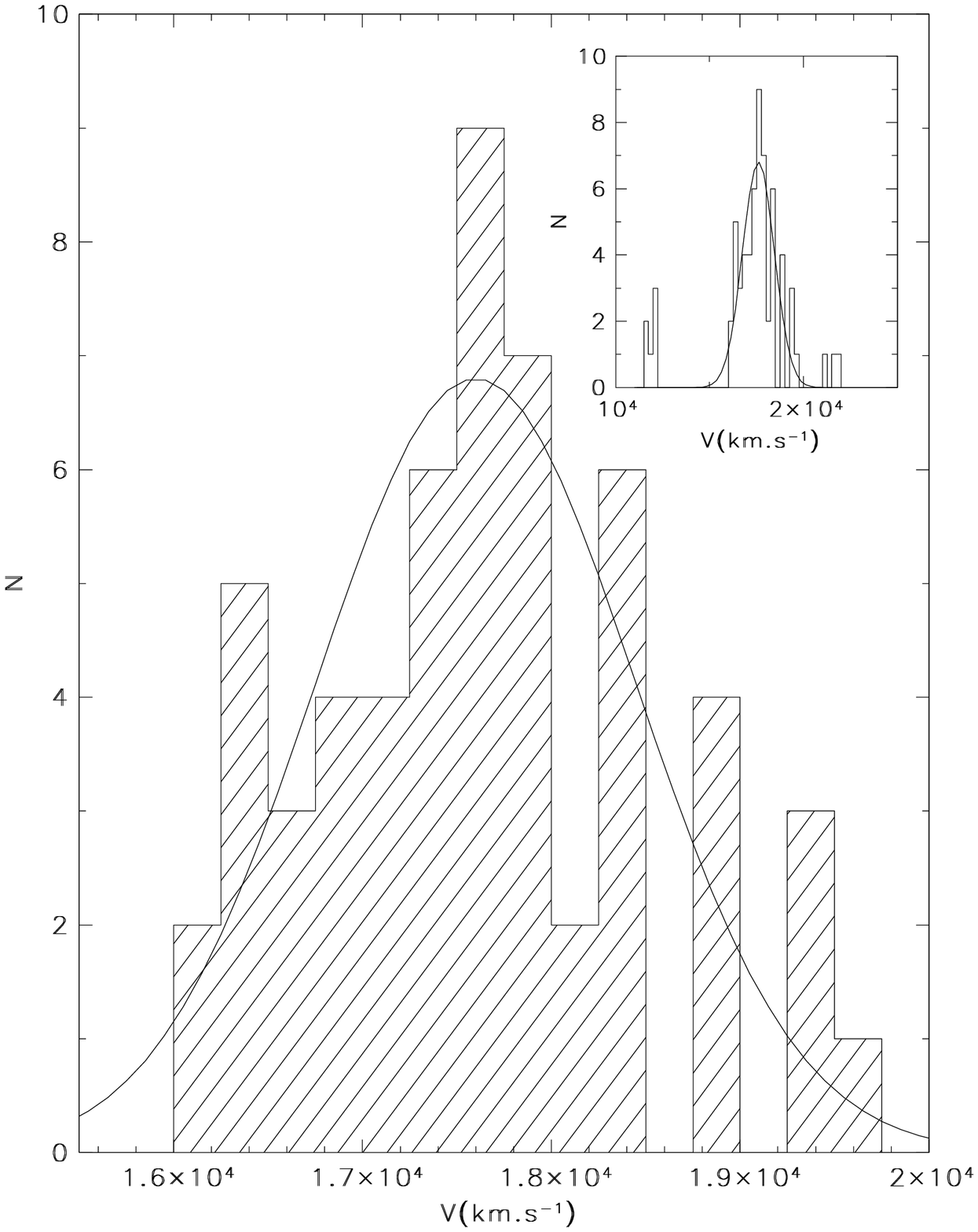}}
\par}
{\centering \small {Fig. 6- The radial velocity distribution for the
Abell 970 sample of galaxies. The continuous curve shows the Gaussian
distribution corresponding to the mean velocity and velocity dispersion 
quoted in the text (normalized to the sample size and range). 
The inset displays the velocity distribution between 10000 and 25000
$\kms$.}}
\vspace{0.3cm}

Considering only the 56 galaxies within this velocity range, the
cluster mean velocity is $\overline{V} = 17600~ \pm~ 118 \kms$
(corresponding to $z = 0.0587$) \footnote{In this paper the mean and
dispersion velocities are given as biweighted estimates, see Beers
et al. (1990).}. For comparison, the radial velocity of the E/D galaxy
located at the center of main cluster  is
$17525~ \pm~ 52$, near that of the whole cluster,
as it should be expected if this is the dominant cluster galaxy. The
cluster velocity dispersion, corrected following Danese, De Zotti \& di
Tullio (1980) is $\sigma_{corr} = 845_{-69}^{+92} \kms$ (at a
confidence level of 68\%).  Fig. 6 presents the radial velocity
distribution of the cluster galaxies, as well as a Gaussian curve with
the same mean velocity and dispersion observed for these galaxies.
Note that this value of $\sigma_{corr}$ is well above the value favoured by
the $\sigma-T_X$ relation, $ \sim 700\kms$ (Sect. 4).

If we consider the morphological types, the mean velocities and
corrected velocity dispersions are:  $\overline{V}= 17631\kms$ and
$\sigma =~846_{-87}^{+125}\kms$ for $E + S0$~ galaxies  (35~objects),
and $\overline{V}=\,17655\kms$ and $\sigma =~841_{-112}^{+185}\kms$ for
$S + I$~galaxies (19~objects). Hence, contrarily to what is observed in
most clusters, where the velocity dispersion of the late type
population tends to be larger than that of the early type population
(Sodr\'e et al. 1989, Stein 1997, Carlberg et al. 1997, Adami et al. 1998), 
in Abell 970 we do not see any significant difference
between the velocity dispersion of these two populations. This might be
another indication -- besides the presence of a substructure -- that 
Abell 970 is not in overall dynamical equilibrium.

\subsection{Substructures in the galaxy distribution}
Let us now consider again the substructure, as well as the peak of the
galaxy distribution (cf. Fig. 2), taking into account the galaxy
velocities. This analysis will be done with galaxies brighter than 
$b_J^{cosmos}=19.0$, the magnitude where the completeness of our 
velocity catalogue in the central regions of the cluster is 75\%.

The substructure at NW of the main cluster has, within a 3 arcmin 
(274 $h_{50}$ kpc) circular region centered on the brightest S0/S
galaxy, 7 cluster galaxies brighter than $b_J^{cosmos}=19.0$.
Together, these galaxies have a low velocity dispersion, $\sigma_{NW} =
378_{-77}^{+120} \kms$, more typical of that of loose groups. The mean
velocity is $\overline{V}_{NW} = 17834~ \pm~ 135\kms$, significantly
higher than the overall mean velocity of the cluster. Our velocity
catalogue contains 2 galaxies fainter that $b_J^{cosmos}=19.0$ inside
this region. Their inclusion does not significantly change
the value of mean velocity, although it increases the velocity dispersion
to $\sigma_{NW} = 525_{-87}^{+160} \kms$, a value, however,
significantly lower than the cluster overall velocity dispersion. 
These results are consistent with the suggestion that this clump of 
galaxies forms a loose group infalling
towards the cluster main central condensation. Arguing
against the reality of such a group, we notice that its dominant S0/S
galaxy is also the lowest velocity member, with $v = 17184\kms$, but this
is not statistically significant.

The central cluster condensation has a N-S elongation (see Fig. 2).
A closer examination of the galaxy distribution indicates that this region is
dominated by two small clumps of galaxies, which we will denote by A and B
(see also Fig. 7 below).
Considering circular
regions of 1 arcmin ($\sim 91 \kpc$), the central clump, A, is tightly
concentrated around the E/D galaxy, having 6 galaxies brighter than
$b_J^{cosmos}=19.0$ with
$\overline{V_A} = 17624\kms$, and dispersion $\sigma_A = 816$.  The
other clump, B, is about 1.5 arcmin NW of clump A and  is more sparse, 
with only 4 galaxies, of which 3 are tightly packed in velocity space
with velocity dispersion $\sigma_B =711\kms$ and mean velocity 
$\overline{V}_B = 19227\kms$. The fourth galaxy that is, apparently,
member of this clump has, however, a very discrepant radial velocity,
$ 16847\kms$. Since it is not apparent in Fig. 2, it is not clear if B 
is a real substructure or a
fortuitous projected group of cluster galaxies.

\subsection{Velocity gradients}
Fig.s 7 and 8 display, respectively, the adaptive kernel maps
for the mean velocity and the mean velocity dispersion
of the sample of cluster galaxies with measured radial
velocities brighter than $b_J^{cosmos}=18.9$.  These 
maps were calculated from the local kernel weighted averages,
with initial kernel size usually larger ~-~ in our case by a factor 3,
as a compromise between signal-to-noise and spatial resolution ~-~ than
the optimal size prescribed by Silverman (1986), as suggested by
Biviano et al. (1996). Significance regions for each map were obtained
by bootstrap, in a fashion similar to that applied to the projected
density maps.

The mean velocity map of Fig. 8 clearly indicates the existence of a
velocity gradient across the field, roughly in the E-W direction. This
occurs because galaxies with $v > 18500\kms$ populate predominantly the
east-side of the field. The mean velocity of the NW group discussed
above is consistent with this gradient. Interestingly, this gradient is
also consistent with the general gradient one would obtain by
considering all galaxies of our catalogue with velocities between
$12000$ and $22000\kms$, suggesting that the cluster may be part of a
larger structure running more or less in the E-W direction within, at
least, this velocity range. In fact a (smaller) velocity gradient, 
in this same general direction, is also depicted by the mean redshifts
of the supercluster members (cf. Einasto et al. 1997).

{\centering
\resizebox*{12cm}{!}{\rotatebox{-90}{\includegraphics{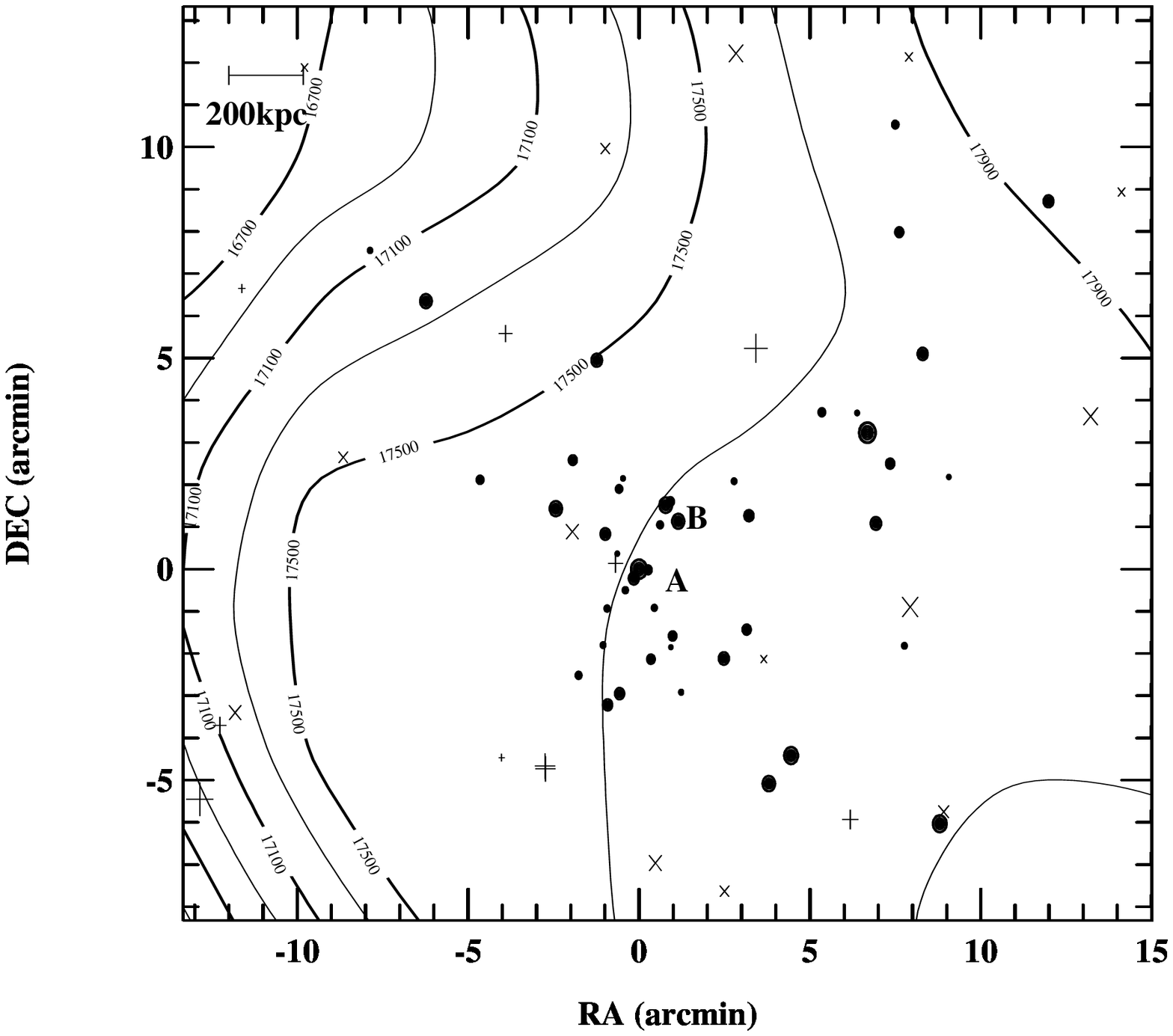}}}
\par}
{\centering \small {Fig. 7 - Mean velocity map of the galaxies
kinematically linked to the cluster and brighter than $b_J^{cosmos}
=18.9$. The positions of COSMOS galaxies brighter than this limit are
also plotted, with symbols following Fig. 3. All regions of this map
are significant. A and B correspond to the galaxy clumps discussed in
Sect. 5.2.}}
\vspace{0.3cm}

The mean velocity dispersion map displayed in Fig. 8 indicates that
there is a radial gradient of the velocity dispersion. 
This was
confirmed by direct calculations of the velocity dispersions within
concentric regions centered in the E/D galaxy. Such a gradient is
expected if the cluster grows through the capture of low velocity
dispersion groups by the central, main galaxy concentration.

The above discussion points to the complexity of the velocity field of
Abell 970. It is possible that the high velocity tail of the cluster velocity
distribution displayed in Fig. 6 may be contaminated by another
component with mean velocity $\overline{V} \gsim 18500\kms$, which
reveals itself by the peculiarities of its spatial distribution. The
fact that the velocity distribution shows some signs of bimodality, as
pointed out at the beginning of this section, reinforces this suggestion.
If real, this component could be interpreted as a diffuse halo
located at the East-side of cluster, probably infalling into its dark
matter potential well. If correct, such a scenario may be shown up by
some X-ray emission features typical of gas shocks produced during this
infall.

{\centering
\resizebox*{12cm}{!}{\rotatebox{-90}{\includegraphics{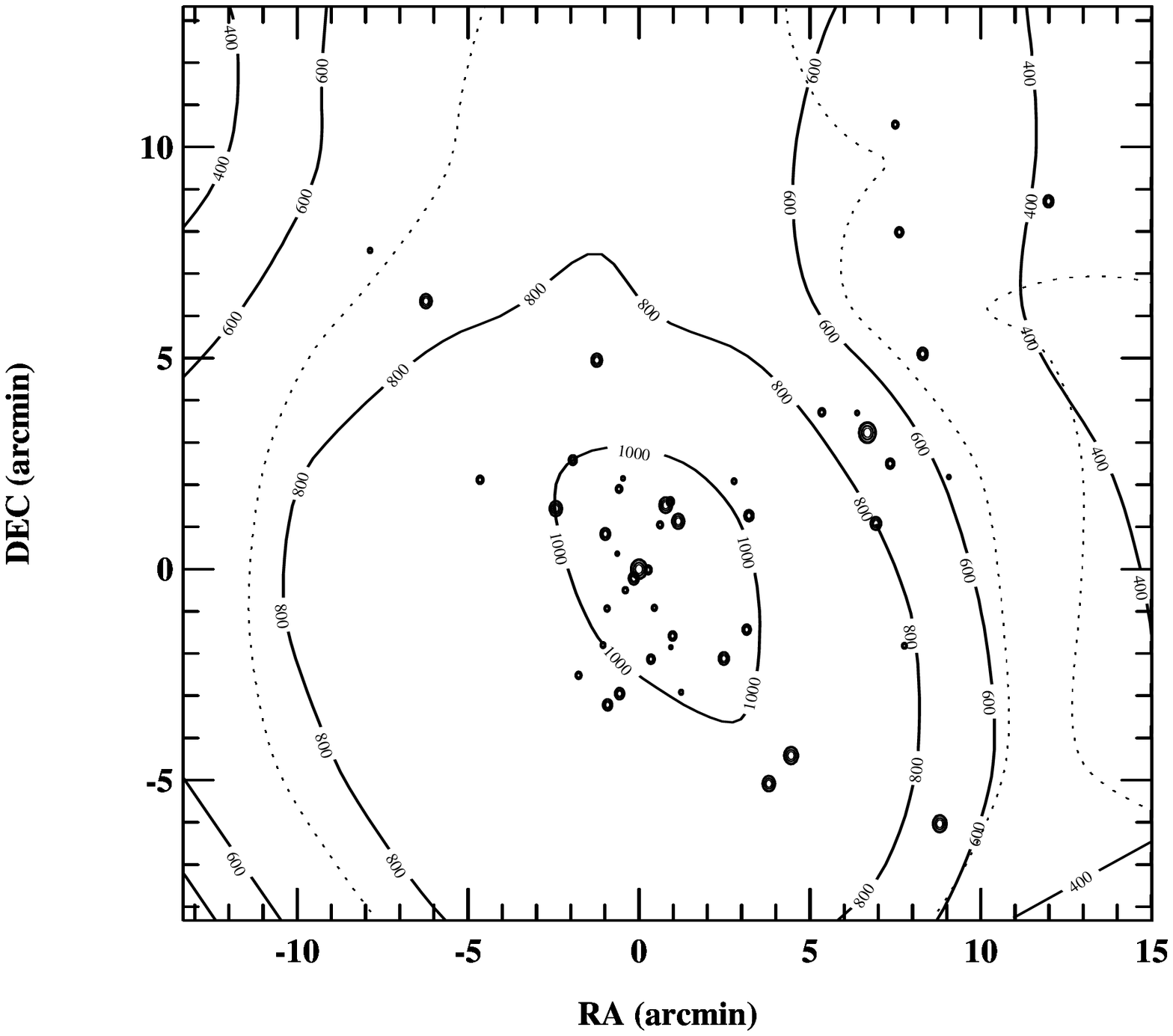}}}
\par}
{\centering \small {Fig. 8 - Mean velocity dispersion map of the galaxies
kinematically linked to the cluster and brighter than $b_J^{cosmos}
=18.9$. The positions of COSMOS galaxies brighter than this limit are
also plotted, with symbols following Fig. 2. The dashed contours delineate
the boundaries of the 99\% significance levels regions of the map.}}
\vspace{0.3cm}

\section{Cluster mass}
In this section we present mass estimates for Abell 970. We first consider
the mass derived from its X-ray emission and then we present masses
computed from the velocity and galaxy distributions.

\subsection{X-ray mass}
Supposing that the gas is isothermal, in hydrostatic equilibrium and
spherically distributed, the dynamical mass inside $r$ is given by:

\begin{equation}
    M(r) = - \frac{k \, T_{X} \, r}{G \, \mu \, m_{p}} \frac{d \log 
    \rho(r)}{d \log r} \, ,
    \label{eq:MassaDinamicaX}
\end{equation}
where $\mu$ is the mean molecular weight ($\mu=0.59$ for a fully ionized
primordial gas) and $m_{p}$ is the proton mass. If we assume that the
gas is described by a $\beta$-model,

\begin{equation}
\rho(r) = \rho_{0} (1 + (r/r_{c})^2)^{-3\beta/2} \, ,
\end{equation}
the dynamical mass becomes:
\begin{equation}
M(r) =
6.68\times 10^{10} \frac{\beta\, T_{X}}{\mu} \frac{r^3}{r^2 + r_{c}^2}~ 
M_{\odot}\, ,
\label{MassaDinamicaNum}
\end{equation}
with $r$ and $r_{c}$ measured in kpc and $T_{X}$ in keV.

Furthermore, if we have the density contrast $\delta =
\bar{\rho}(r)/\rho_{c}$ (where $\bar{\rho}(r)$ is the mean density
inside the radius $r$ and $\rho_{c}$ is the critical density at the
redshift of the cluster) then we can define $r_{\delta}$ as
\begin{equation}
\left(\frac{r_{\delta}}{r_{c}}\right)^2 = \frac{5.75\times 10^7}{\delta \, h^2\,
f^2(z, \Omega_{0}, \lambda_{0})} \frac{\beta\, T_{X}}{\mu \, r_{c}^2} \, ,
\end{equation}
where $f(z, \Omega_{0}, \lambda_{0})$ depends on the cosmological parameters:
\begin{equation}
f^2(z, \Omega_{0}, \lambda_{0}) = \lambda_{0} + \Omega_{0} (1+z)^3 
-(\Omega_{0}+\lambda_{0}) (1+z)^2 \, .
\end{equation}
When $\delta=200$ we have the usual $r_{200}$ radius, which is about 
the virial radius, $r_{\rm vir}$ (e.g. Lacey \& Cole 1993).

In order to apply the above formulae, we need the parameters for the
$\beta$-model that describe the intra-cluster gas. We estimate these
parameters using the correlations presented by Jones \& Forman (1999)
between both $\beta$ and $r_{c}$ with $T_{X}$ (Figs. 6 and 7 in their
paper). Taking $T_{X}=3.1$~keV, we have $\beta \approx 0.5$ and $r_{c}
\approx 100$~kpc.  Therefore, we obtain $r_{200}/r_{c} = 16.2$ or
$r_{200} = 1620$~kpc. The corresponding mass is $M(r_{200}) =
2.84\times 10^{14} M_{\odot}$. This result agrees well with the virial
mass--temperature of Horner, Mushotzky \& Scharf 1999 (cf. their
Fig.~1 and Table~1).  The total mass in the region where the
velocities have been measured is $M(r=1.2 \rm h_{50}^{-1} Mpc) = 2.07
\times 10^{14} M_{\odot}$.

\subsection{Optical Virial Mass}
In Fig. 9 we show the cluster mass profile computed with the
virial mass estimator (VME) which, as discussed by Aceves \& Perea (1999), 
gives the less biased mass estimates when the system
is not completely sampled. These authors also show that the VME 
overestimates the real mass by no more than 20\% at small radii, being more 
reliable at larger apertures. The error bars in Fig. 9 are 1-$\sigma$
standard deviations computed using the bootstrap method. The VME assumes,
of course, that the system is virialized. In general, the presence of
substructures or large scale flows tend to increase the velocity 
dispersion of the galaxies, leading to an overestimation of the mass 
of the system.

The VME of Abell 970, within 1.2 $h_{50}^{-1}$ Mpc, is 
$M = (6.80 \pm 1.34)
\times 10^{14} M_{\odot}$, where the error, as before, was computed with 
the bootstrap method.  Note that, for a virialized cluster, these
are lower limits for the mass, since we have velocities only for the
central region of the system.  Indeed, assuming a relation between
virial radius and velocity dispersion similar to that adopted by
Girardi et al. (1998), we estimate that $r_{vir} \sim 3.4 h_{50}^{-1}$
Mpc, while the velocities have been measured within a region of radius
$\sim 1.2 h_{50}^{-1}$ Mpc.

Fig. 9 also displays the run of the total $b_j$ luminosity of the
cluster (up to $b_j = 19.75$).
Considering the VME masses, we find that the the mass-luminosity ratio
range from 1360 at the cluster central region, to $\sim$ 450 at the
largest aperture.

The mass profile derived from the X-ray emission is also presented in
Fig. 9.
The VME masses are in excess to the X-ray mass estimates
by large factors, ranging from $\sim$ 16 for the central apertures, to
about 4 at $\sim$ 1.3 $h_{50}^{-1}$ Mpc aperture, which encompasses the
whole velocity sample. These factors are well above the uncertainties
discussed above for virialized clusters. In fact,
the dynamical mass determined by the X-ray emission at radius
$r \gg r_c$ depends essentially on the temperature and the
asymptotical slope of the gas density. Both are ill determined
with the available data; it is then possible that one of them
(or both) are under-estimated, which implies that we under-estimate
the dynamical X-ray mass. For instance, if $\beta$ is as high as
0.75 and $T_X = 4.1$ (cf. the error bars in Table~2), then the dynamical 
X-ray mass would be twice the estimated value, i.e., 
$M(r=1.2$h$_{50}^{-1}$Mpc$)=4.1\times10^{14}M_\odot$.
On the other side, the presence of a substructure associated to the
cluster brightest galaxy, as well as the mean velocity gradient, may
be an indication of non-virialization and, consequently, the VME
may be largely overestimated.

\vspace{0.3cm}
{\centering
\resizebox*{12cm}{!}{\rotatebox{0}{\includegraphics{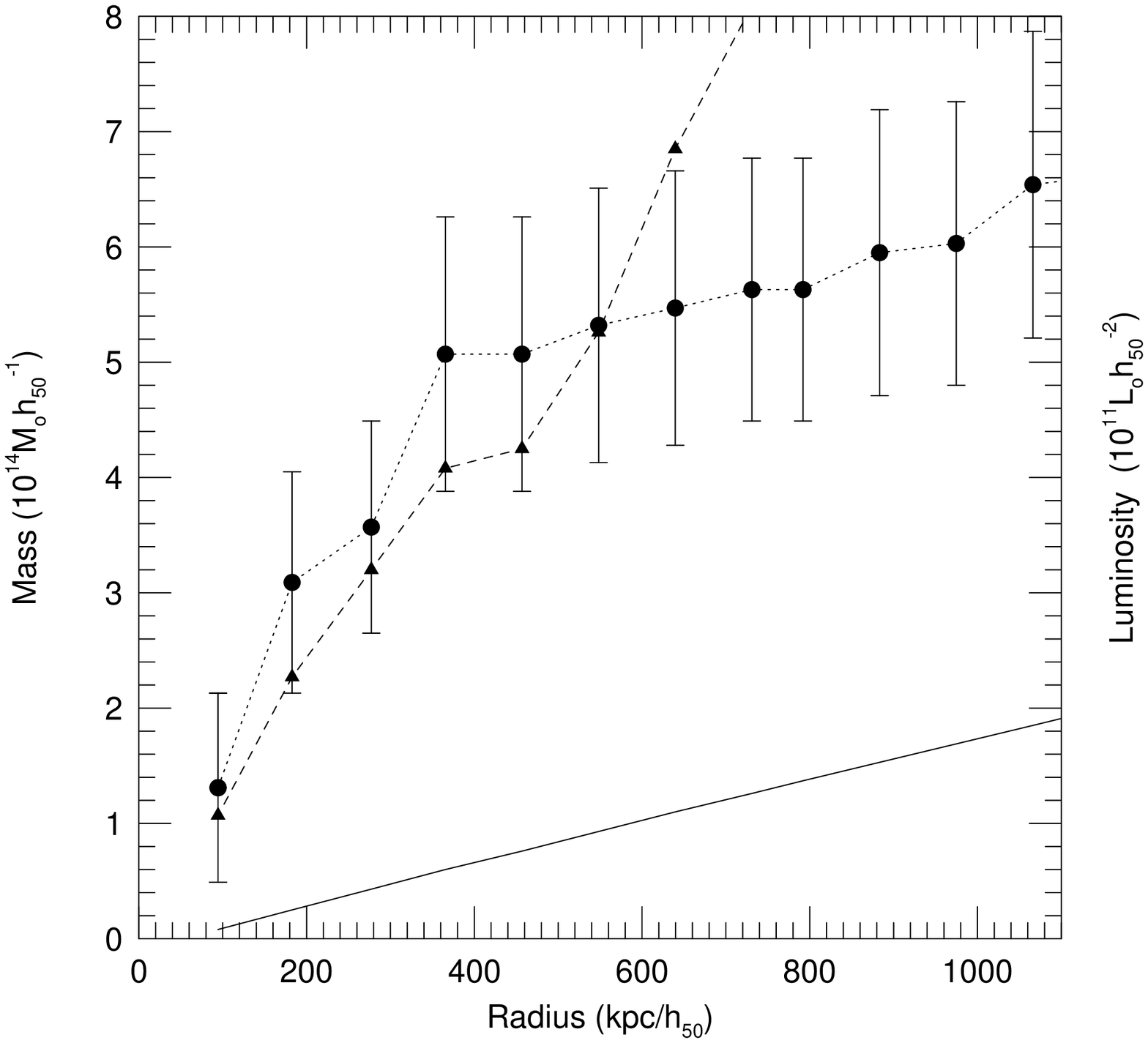}}}
\par}
{\centering \small {Fig. 9 - Cluster optical masses (circles and
dotted lines) and luminosities (triangles and dashed lines). The X-ray
masses are given by the continuous line. Filled symbols give estimates
using the whole sample of velocities.}}
\vspace{0.3cm}

\section{The dynamical status of Abell 970}
Several evidences indicate that Abell 970 is not in an overall state
of virialized equilibrium. Indeed, the presence of a
substructure at NW of the main galaxy concentration
may indicate that the cluster has been capturing groups of 
galaxies of its neighborhood. The large scale gradient 
in the mean velocity of the galaxies,
shown in Fig. 8, is also not expected in virialized systems.
Other evidences for non-equilibrium
include the discrepance in the peaks of the X-ray emission and of
the galaxy projected density, and the observed differences between
optical and X-ray masses.

Accordingly to Allen (1998), there is a good agreement between
X-ray and strong gravitational lensing mass measurements only in
clusters with strong cooling flows; in clusters with modest or absent cooling
flows, the masses determined from the X-ray data are 2 to 4 times 
smaller than the masses estimated from strong gravitational lensing.
The reason for the mass dicrepancy is the dynamical status of the central
regions of the clusters: those with strong cooling flows are relaxed and
virialized, while those with small cooling flows are out of the equilibrium,
and the assumption of hydrostatic equilibrium that underlies the X-ray mass
estimates is not valid. Indeed, the presence or absence of cooling flows can
be used as a diagnostic to verify whether galaxy clusters are in dynamical 
equilibrium in their central regions.
Abell 970 has at most a weak cooling flow,
of $\dot{M} = 20^{+32}_{-20}M_\odot$ yr$^{-1}$ (Ebeling et al. 1996), and
our results, indicating that this system is not relaxed, 
are consistent with the findings of Allen (1998). 

It is interesting to note that the offset between the X-ray and galaxy
distribution centers, although significant, is not very large. It is
possible that Abell 970 had a much stronger cooling flow until recently, that
was interrupted by dynamical perturbations induced by the arrival of a
galaxy group (now observed as a substructure) in
the central regions of the cluster.
Given that in the cluster densest regions the gas relaxes very quickly, 
compared with the time the galaxy distribution takes to achieve equilibrium,
it is natural to think of cooling flows as an intermittent process, that 
is disrupted by dynamical perturbations and that resumes 
activity after the relaxation is achieved. The time scale of intermittency,
in this scenario, depends strongly on the accretion rate of groups by the 
cluster.

\section{Summary}

In this paper, we have presented an analysis of the galaxy cluster Abell~970,
based on a new set of radial velocities and on X-ray observations. The study of
the galaxy projected positions reveals a relatively regular distribution,
centered on an E/D dominant galaxy. The analysis with the adaptive kernel
density map indicates the presence of a statistically significant substructure
at NW of the cluster main galaxy concentration, centered on a S0/S galaxy that
is the brightest cluster member. The X-ray emission distribution does not reveal
any emission excess due to that substructure but, interestingly, the peak of the
X-ray emission is not coincident with the cluster center (at the position of the
dominant galaxy), being displaced towards the direction of the substructure.
These results suggest that this substructure is real and that the cluster may
not be in an overall state of dynamical equilibrium.

Further evidence that the cluster is in a state of non-equilibrium comes
from the analysis of the radial velocity distribution. For instance,
the cluster velocity dispersion, 845 km s$^{-1}$
(increasing to $\sim$1000 km s$^{-1}$ at the cluster center), 
is significantly larger
than the value expected from the $\sigma - T_X$ relation, that is 
$\sim$700 km s$^{-1}$. Also, the substructure detected in the galaxy projected
distribution has a much smaller velocity dispersion, 381 km s$^{-1}$, that is
typical of loose groups. Together, these results suggest that this
substructure may be a group that have recently arrived in the central
regions of the cluster. The presence of large scale velocity gradients
is another evidence that Abell 970 is out of equilibrium.
The virial mass of this cluster is much larger than the mass inferred from
the X-ray emission. This discrepancy is indeed expected if the underlying
hypothesis of these mass estimators, namely that galaxies and 
gas inside the cluster are in hydrostatic equilibrium, is not actually
fullfiled.

The fact that Abell 970 has a dim cooling flow also fits nicely in the above 
scenario if, as suggested by Allen (1998), only clusters in equilibrium exibit
massive cooling flows. Indeed, cooling flows may have an intermittent behavior:
phases of massive cooling flows may be followed by phases without significant
cooling flows after the accretion of a galaxy group massive enough to 
disrupt the dynamical equilibrium in the center of the clusters. 
After a new equilibrium is
achieved, a massive cooling flow will be established again. 
Hence, in hierarchical scenarios for structure formation, 
intermittent cooling-flows should be an usual phenomenon.

\acknowledgements {We thank the OHP, Pic du Midi and ESO staff for their
assistance during the observations, and especially Gilles Charvin, student
at the Ecole Normale Sup\'erieure de Lyon for his valuable scientific
collaboration. LSJ, HVC, GBLN and HC thank the financial support provided 
by FAPESP, CNPq and PRONEX.
DP acknowledges support from ECOS/CONICYT project C96U04.
This research has made use of data obtained through the High 
Energy Astrophysics Science Archive Research Center Online Service, 
provided by the 
NASA/Goddard Space Flight Center.}


\end{document}